\documentclass[aps,prl,reprint,superscriptaddress,nofootinbib]{revtex4-2}

\usepackage{amsmath,amssymb,bm}
\usepackage{graphicx}
\usepackage{booktabs}
\usepackage{siunitx}
\usepackage{microtype}
\usepackage[colorlinks=true,citecolor=blue,linkcolor=blue,urlcolor=blue]{hyperref}

\newcommand{\ga}{g_{a\gamma}}
\newcommand{\ma}{m_a}
\newcommand{\Pag}{P_{a\to\gamma}}
\newcommand{\Bperp}{\mathbf B_{\perp}}
\newcommand{\IB}{\mathcal I_B}
\newcommand{\EB}{\mathcal E_B}

\begin{document}

\title{Light Shining Through the Moon: A Lunar-Occultation Helioscope for Solar Axions}

\author{D. Suárez-Fontanella}
\email{duvier@usal.es}

\affiliation{Department of Fundamental Physics and IUFFyM, University of Salamanca,
	Plaza de la Merced SN, E–37008 Salamanca, Spain}

\date{\today}

\begin{abstract}
We propose a lunar-occultation helioscope in which the Moon suppresses the direct solar x-ray beam before solar axions enter the terrestrial magnetosphere, where they may reconvert into keV photons. We evaluate the conversion with a geomagnetic model combining the IGRF-14 internal contribution and the Tsyganenko T89 external magnetospheric contribution, retaining the two signed transverse components in the oscillation amplitude. As an explicit realizability test, we propagate a circular near-600-km polar orbit through the 12 August 2026 occultation. A $4'$-radius region around the solar center remains lunar occulted for about $470$ s while the sunward line of sight stays clear of the Earth; the coherent magnetic lever arm spans $69.1$--$96.4~\mathrm{T\,m}$ with $\mathcal I_{B,\mathrm{rms}}=76.9~\mathrm{T\,m}$. A B16-AGSS09 Primakoff calculation shows that this aperture contains $97.4\%$ of the 3--15 keV axion flux. For an explicitly specified NuSTAR-like detector benchmark, the propagated single-event exposure gives an expected low-mass 95\% C.L. sensitivity of $\ga=1.31\times10^{-10}~\mathrm{GeV}^{-1}$. The proposal therefore establishes the plausibility of a distinct experimental channel for solar-axion searches, characterized by a lunar-defined spatial and temporal window and by a predictable geomagnetic modulation. A future detection would additionally provide directional access to the axion-emitting solar core, opening a complementary probe of the solar interior.
\end{abstract}

\maketitle

Axions arise from the Peccei--Quinn solution of the strong-$CP$ problem \cite{PecceiQuinn1977,Weinberg1978,Wilczek1978}, while axionlike particles (ALPs) occur more generally in extensions of the Standard Model. Their two-photon interaction,
\begin{equation}
\mathcal L_{a\gamma}=-\frac{\ga}{4}aF_{\mu\nu}\widetilde F^{\mu\nu} = \ga a\,\mathbf E\!\cdot\!\mathbf B ,
\label{eq:Lint}
\end{equation}
allows thermal photons in the solar plasma to produce axions through the Primakoff process and permits the inverse conversion in an external magnetic field \cite{Sikivie1983,RaffeltStodolsky1988}. CAST currently sets $\ga<5.8\times10^{-11}~\mathrm{GeV}^{-1}$ at 95\% C.L. for $\ma\lesssim0.02~\mathrm{eV}$ \cite{CAST2024}, while BabyIAXO and IAXO target substantially improved helioscope sensitivity \cite{BabyIAXO2021}.

The Earth's magnetic field has long been recognized as a possible conversion region. Davoudiasl and Huber proposed geomagnetic conversion of solar axions in low Earth orbit \cite{DavoudiaslHuber2006} and subsequently developed a realistic orbital feasibility study (GECOSAX) including a nonuniform geomagnetic field, x-ray absorption in the atmosphere, and measured satellite backgrounds \cite{DavoudiaslHuber2008}. More recently, NuSTAR observations have constrained axions converting in the solar atmospheric magnetic field, reaching $\ga\lesssim7.3\times10^{-12}~\mathrm{GeV}^{-1}$ for $\ma\lesssim4\times10^{-7}~\mathrm{eV}$ \cite{Ruz2025}. Solar-eclipse data have also been used in dark-photon searches, where the Moon--Earth distance can itself become an oscillation baseline \cite{Frerick2023}.

\begin{figure*}[t]
\centering
\includegraphics[width=0.93\textwidth]{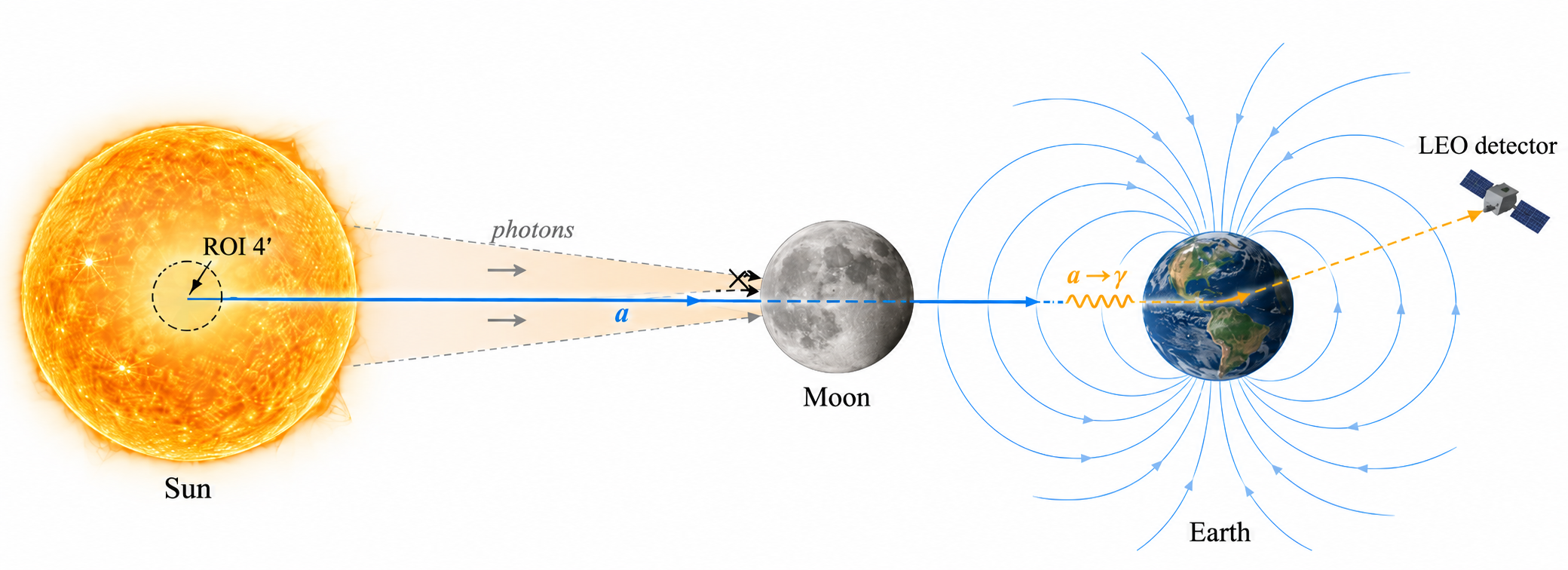}
\caption{Lunar-occultation helioscope. Solar photons are absorbed by the Moon, whereas solar axions traverse the lunar body and continue into the terrestrial magnetosphere. A transverse geomagnetic field can reconvert a fraction of the axion flux into keV photons before it reaches an orbital x-ray detector. The illustration is schematic and not to scale.}
\label{fig:concept}
\end{figure*}

Here the Moon plays a different role, it is an \emph{upstream occultor}, direct solar x rays are removed before the weakly interacting beam reaches the terrestrial magnetosphere, whereas any regenerated photons are produced downstream of the Moon. The arrangement, shown schematically in Fig.~\ref{fig:concept}, is therefore analogous to a light-shining-through-a-wall experiment \cite{vanBibber1987}, the Sun as source, the Moon as photon wall, the geomagnetic field as regeneration region, and an orbital x-ray instrument as detector.

\textit{Geomagnetic regeneration.---}
For an axion of energy $E$ propagating along a solar line of sight with unit vector $\hat{\ell}$, define
\begin{equation}
\Bperp(s)=\mathbf B(s)
-\left[\mathbf B(s)\!\cdot\!\hat{\ell}\right]\hat{\ell}.
\label{eq:Bperp}
\end{equation}
We introduce a fixed orthonormal basis $\{\hat{\mathbf e}_1,\hat{\mathbf e}_2\}$ in the plane perpendicular to $\hat{\ell}$ and retain the signed components
\begin{equation}
B_i(s)=\mathbf B(s)\!\cdot\!\hat{\mathbf e}_i,\qquad i=1,2.
\label{eq:Bcomponents}
\end{equation}
In a slowly varying magnetized plasma the two photon-polarization amplitudes are
\begin{equation}
\mathcal A_i(E,\ma) =\frac{\ga}{2}\int ds\,B_i(s) \exp\!\left[ i\int_0^s ds'\,\frac{\ma^2-\omega_{\rm pl}^2(s')}{2E}\right],
\label{eq:amplitude}
\end{equation}
and
\begin{equation}
\Pag(E,\ma)=|\mathcal A_1|^2+|\mathcal A_2|^2 .
\label{eq:probability}
\end{equation}
The plasma frequency is
\begin{equation}
\omega_{\rm pl}\simeq 3.69\times10^{-11} \sqrt{\frac{n_e}{\mathrm{cm}^{-3}}}\ {\rm eV}.
\label{eq:plasma}
\end{equation}
For the plasma contribution we adopt a smooth semi-empirical electron-density profile whose inner plasmaspheric branch and $K_p$-dependent plasmapause are based on the Carpenter--Anderson model \cite{CarpenterAnderson1992}, with a smooth $r^{-2}$ continuation into the outer magnetosphere; the full prescription is given in the Supplemental Material. For the internal field we use IGRF-14 \cite{Beggan2026} and T89 for the large-scale external magnetosphere \cite{Tsyganenko1989}. The magnetic profile is always evaluated along the \emph{solar} line of sight; the Moon enters only through the occultation condition.

In the coherent limit,
\begin{equation}
\IB\equiv\left[\left(\int B_1 ds\right)^2+\left(\int B_2 ds\right)^2\right]^{1/2},\,
\Pag^{\rm coh}=\frac{\ga^2}{4}\IB^2 .
\label{eq:IB}
\end{equation}
Thus $\IB$ is the magnitude of the coherently integrated transverse vector. With the units used below, $1~\mathrm{T\,m}=0.9882~\mathrm{GeV}$ in natural units, providing a direct normalization check on Eq.~(\ref{eq:IB}).

For the full-Sun Primakoff spectrum we use \cite{CAST2024}
\begin{equation}
\frac{d\Phi_a}{dE} = 6.02\times10^{10}g_{10}^{\,2}E^{2.481}e^{-E/1.205}\ \mathrm{cm}^{-2}\mathrm{s}^{-1}\mathrm{keV}^{-1},
\label{eq:flux}
\end{equation}
where $E$ is in keV and $g_{10}=\ga/(10^{-10}\,\mathrm{GeV}^{-1})$. The analysis aperture is a circular region of angular radius
\begin{equation}
\theta_{\rm ROI}=4.0',
\label{eq:roi}
\end{equation}
centered on the apparent solar center. At the fiducial epoch the solar angular radius is $15.7779'$, corresponding to $\rho_{\rm ROI}\equiv\theta_{\rm ROI}/\alpha_\odot=0.253519$. The signal therefore uses
\begin{equation}
\frac{d\Phi_a^{\rm ROI}}{dE} =f_{\rm ROI}(E)\frac{d\Phi_a}{dE},
\label{eq:roiflux}
\end{equation}
where $f_{\rm ROI}$ is computed from the projected Primakoff image of the B16-AGSS09 standard solar model \cite{Vinyoles2017,Hoof2023}. Over 3--15 keV the spectrum-weighted fraction is $0.973734$; for example, $f_{\rm ROI}=0.9196$ at 3 keV and $0.9828$ at 5 keV. The $4'$ aperture is therefore a compact choice that retains $97.4\%$ of the 3--15 keV Primakoff signal while covering only $\simeq6.4\%$ of the apparent solar-disk area, thereby limiting the extraction region without significant loss of signal.

\textit{Propagated lunar-occultation orbit.---}
Complete occultation of the adopted signal region requires
\begin{equation}
\delta_{\odot M}(t)+\theta_{\rm ROI}<\alpha_M(t),
\label{eq:occult}
\end{equation}
where $\delta_{\odot M}$ is the apparent Sun--Moon center separation and $\alpha_M$ is the apparent lunar radius. We construct a circular polar orbit passing through the fiducial 12 August 2026 occultation position at 17:45:54 UTC and propagate it with two-body dynamics. Its geocentric radius is $6961.845~\mathrm{km}$, its speed $7.5667~\mathrm{km\,s^{-1}}$, and its period $96.349~\mathrm{min}$. The corresponding anchor position is at a nominal WGS84 altitude of 600 km.

\begin{figure}
    \centering
    \includegraphics[width=0.5\linewidth]{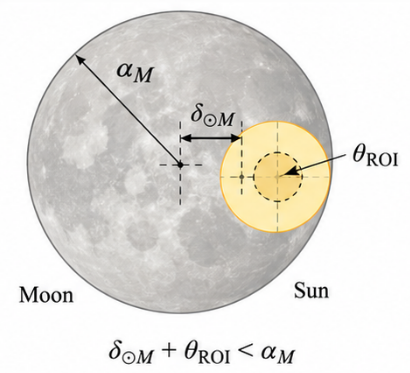}
    \caption{Lunar-occultation geometry defining the axion search region. During solar occultation, the lunar disk blocks direct photons from the selected solar ROI, whereas the axion flux continues along the original Sun–observer direction. The region projected behind the Moon therefore provides a low-background window in which photons regenerated through $a\rightarrow\gamma$ conversion along the Earth-directed trajectory can be searched for.}
    \label{fig:placeholder}
\end{figure}

The ROI remains lunar occulted while the sunward line of sight stays clear of the Earth from $t-t_{\rm occ}=-250$ to $+220$ s, giving
\begin{equation}
\Delta t_{\rm occ}\simeq470~{\rm s}.
\end{equation}
Across this interval
\begin{equation}
69.147\le \IB(t)\le96.420~{\rm T\,m}.
\end{equation}
Because the coherent rate scales as $\IB^2(t)$, we define
\begin{equation}
\EB\equiv\int_{\rm occ}dt\,\IB^2(t) =2.778552\times10^6~({\rm T\,m})^2\,{\rm s},
\label{eq:magnetic_exposure}
\end{equation}
and
\begin{equation}
\mathcal{I}_{B,\mathrm{rms}} =\left(\frac{\EB}{\Delta t_{\rm occ}}\right)^{1/2} =76.888~{\rm T\,m}.
\label{eq:IBrms}
\end{equation}
The propagated state reproduces the independently evaluated anchor $\IB$ to $0.022\%$, providing a numerical geometry check.

\begin{figure}[t]
\centering
\includegraphics[width=\columnwidth]{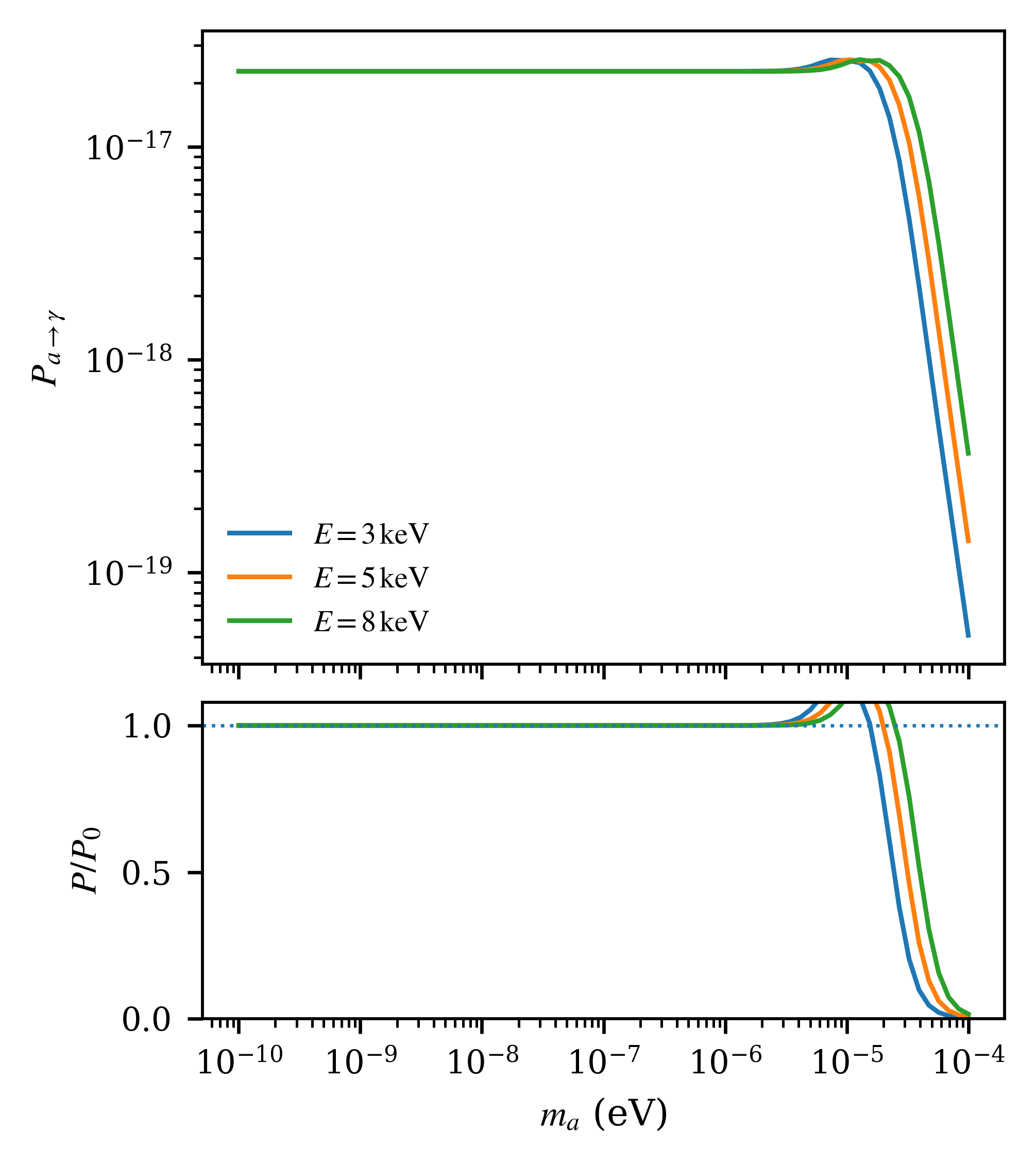}
\caption{Phase-resolved geomagnetic conversion along the propagated August orbit at the occulted sample with the largest coherent lever arm. Top: $\Pag(\ma)$ for $E=3$, 5, and 8 keV at $\ga=10^{-10}~\mathrm{GeV}^{-1}$. Bottom: ratio to the corresponding low-mass plateau, $P/P_0$. Both signed transverse components are propagated and the unresolved final photon polarizations are summed.}
\label{fig:coherence}
\end{figure}

Figure~\ref{fig:coherence} evaluates the full phase-resolved probability at the occulted orbital sample with the largest $\IB$. The low-mass probability is of order $10^{-17}$ for $\ga=10^{-10}~\mathrm{GeV}^{-1}$, and coherence is progressively lost above the few-$10^{-5}$ eV scale. Refining the maximum phase step from $0.10$ to $0.05$ rad changes the probability by only $0.125\%$ at $\ma=10^{-4}$ eV.

\textit{Single-event sensitivity benchmark.---}
The expected number of regenerated photons is obtained by integrating over the physical occultation window and the 3--15 keV analysis band,
\begin{align}\label{eq:signal}
S(\ma,\ga)= \int_{\rm occ} dt \int_{3\,{\rm keV}}^{15\,{\rm keV}} dE\, A_{\rm eff}(E)\, \frac{d\Phi_a^{\rm ROI}}{dE}\, \\
\times\Pag(E,\ma,t).\nonumber
\end{align}
For the detector benchmark we adopt a smooth NuSTAR-like focusing-telescope effective area and a total 3--15 keV background rate $R_{\rm bkg}=2.0\times10^{-3}~\mathrm{count\,s^{-1}}$ within the $4'$ ROI. The explicit effective-area points and background prescription are given in the Supplemental Material.

At the reference coupling $\ga^{\rm ref}=10^{-10}~\mathrm{GeV}^{-1}$, the low-mass limit of Eq.~(\ref{eq:signal}) gives
\begin{equation}
S_{\rm ref}\equiv S(\ma\rightarrow0,\ga^{\rm ref})=0.8717
\end{equation}
expected signal counts. The corresponding expected background over the $470$-s occultation is
\begin{equation}
B=R_{\rm bkg}\Delta t_{\rm occ}=0.9400
\end{equation}
counts.

The Primakoff solar flux scales as $\ga^2$, while geomagnetic conversion contributes an additional factor $\ga^2$. The expected signal therefore scales as
\begin{equation}
S(\ma,\ga)= S_{\rm ref}(\ma) \left(\frac{\ga}{\ga^{\rm ref}}\right)^4 ,
\end{equation}
where $S_{\rm ref}(\ma)\equiv S(\ma,\ga^{\rm ref})$.

For the background-only Asimov data set we use
\begin{equation}
q_A= 2\left[ S-B\ln\!\left(1+\frac{S}{B}\right) \right],
\label{eq:asimov}
\end{equation}
and define the expected one-sided 95\% C.L. benchmark by $q_A=2.71$. In the low-mass plateau this gives
\begin{equation}
\ga^{95}=1.314\times10^{-10}~\mathrm{GeV}^{-1}.
\end{equation}
The expected reach remains nearly constant up to $\ma\sim10^{-5}~\mathrm{eV}$ and subsequently weakens as coherence is lost, reaching $4.00\times10^{-10}~\mathrm{GeV}^{-1}$ at $\ma=10^{-4}~\mathrm{eV}$.

\begin{figure}[t]
\centering
\includegraphics[width=\columnwidth]{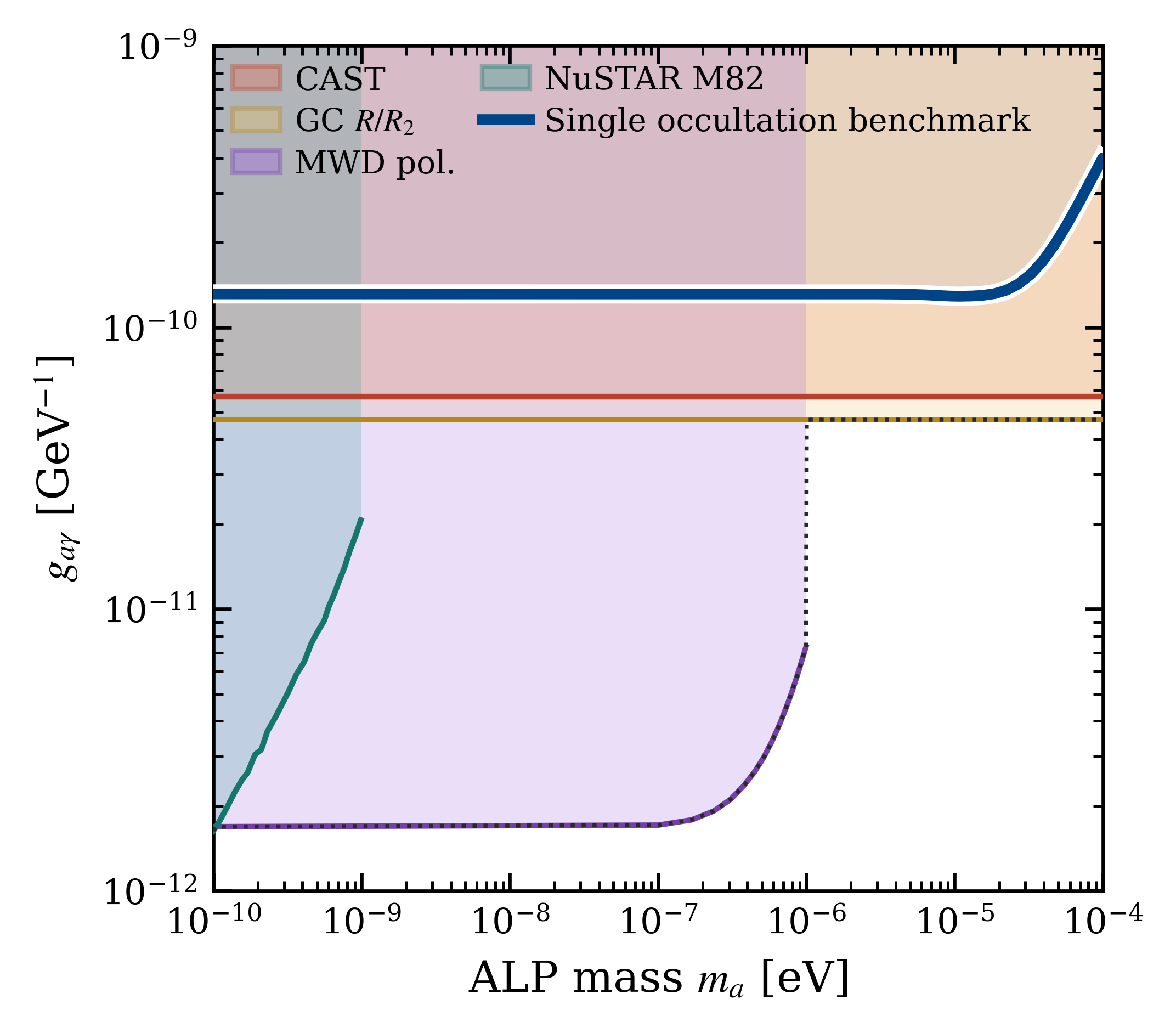}
\caption{Expected 95\% C.L. sensitivity to the axion-photon coupling for the propagated 12 August 2026 lunar-occultation benchmark, compared with existing constraints from CAST, globular-cluster stellar evolution, magnetic-white-dwarf polarization, and NuSTAR observations of M82. The blue curve corresponds to a single, non-optimized $470\,\mathrm{s}$ occultation window. Because the signal scales with the accumulated magnetic exposure, $S\propto \int dt\,\mathcal I_B^2(t)$, magnetically favorable geometries and repeated occultations shift the projected sensitivity toward smaller $g_{a\gamma}$.}
\label{fig:sensitivity}
\end{figure}

\textit{Optimization headroom and accumulated exposure.---}
The sensitivity shown in Fig.~\ref{fig:sensitivity} a single-occultation benchmark, the propagated orbit was selected only to demonstrate dynamical realizability: its altitude, orbital plane, occultation epoch, and orientation of the Sun--Moon--spacecraft line of sight relative to the geomagnetic field were not optimized for axion conversion. In the coherent regime, the relevant quantity is not the peak field integral alone but the magnetic exposure
\begin{equation}
    \mathcal E_B = \int_{\rm occ} dt\,\mathcal I_B^2(t),
\end{equation}
so that, for fixed detector response and solar model,
\begin{equation}
    S_{\rm ref}\propto A_{\rm eff}\,\mathcal E_B .
    \label{eq:signal_scaling}
\end{equation}
Consequently, both the selection of magnetically favorable occultations and the accumulation of several events can substantially increase the expected signal without changing the underlying detection principle.

The amount of available headroom can be illustrated using the most favorable magnetic configuration already reached along the propagated reference orbit. If an orbit were arranged such that an effective $\mathcal I_{B,\rm rms}\simeq96.4~{\rm T\,m}$ were maintained throughout a $470$-s occultation, the magnetic exposure would increase relative to the August benchmark by
\begin{equation}
\frac{\mathcal E_B^{\rm opt}}{\mathcal E_B^{\rm Aug}} \simeq \left(\frac{96.4}{76.9}\right)^2 \simeq1.57 .
\end{equation}
At the reference coupling this would raise the expected low-mass signal from $S_{\rm ref}=0.872$ to approximately $1.37$ counts per occultation. Retaining the conservative benchmark background of $0.94$ counts per $470$-s window, the same Asimov construction gives $\ga^{95}\simeq1.17\times10^{-10}~{\rm GeV}^{-1}$ for one favorable occultation, $8.13\times10^{-11}~{\rm GeV}^{-1}$ after ten equivalent windows, and $5.80\times10^{-11}~{\rm GeV}^{-1}$ after approximately $117$ windows. The latter corresponds to $\simeq15.3$ h of accumulated occulted exposure and reaches the current CAST scale.

The lunar geometry is potentially still more advantageous if the residual background during occultation can be driven well below one count per event. In the ideal background-free Poisson limit, a null observation corresponds to an upper signal yield of approximately three counts at $95\%$ C.L. Under the same favorable magnetic-exposure benchmark, ten accumulated $470$-s occultations would then give $\ga^{95}\simeq6.84\times10^{-11}~{\rm GeV}^{-1}$, while twenty would reach $\simeq5.75\times10^{-11}~{\rm GeV}^{-1}$, slightly beyond the present CAST bound, with only $\simeq2.6$ h of total occulted exposure.

The result may be placed in context with the earlier GECOSAX feasibility study, which assumed a detector area of $10^3~\mathrm{cm^2}$ and roughly $10^6$ s of usable exposure, projecting a $2\sigma$ sensitivity $\ga<(4.7\text{--}6.6)\times10^{-11}~\mathrm{GeV}^{-1}$ for $\ma<10^{-4}$ eV \cite{DavoudiaslHuber2008}. The present calculation addresses a different observing regime, namely a single, short lunar-occultation window, and is therefore not intended as a like-for-like exposure comparison. Its distinctive feature is that the Moon suppresses the direct solar x-ray beam \emph{upstream} of geomagnetic regeneration, providing a geometrically defined dark field together with an externally predicted temporal window and magnetic modulation. Unlike Earth-shadow configurations, lunar occultation decouples solar-photon suppression from the spacecraft position, allowing magnetically favorable, Earth-clear sunward lines of sight, albeit with a smaller and more episodic observing duty cycle.

This structure yields several correlated signal handles. A candidate event must occur at the occulted solar position, have the expected few-keV spectrum, appear only while the selected ROI lies behind the lunar disk, and vary with the calculable transverse geomagnetic field along the orbit. The concept also requires no dedicated conversion magnet or magnet bore. The practical challenge is instead orbital x-ray instrumentation capable of pointing close to the occulted Sun while satisfying field-of-regard, thermal, stray-light, and background constraints.

The circular $600$-km orbit considered here is only one realization of the occultation geometry. Elliptical orbits provide additional freedom to optimize the magnetic exposure: placing an occultation near apogee can increase the dwell time within the lunar shadow, while the orbital orientation can be chosen to favor the transverse geomagnetic field along the solar line of sight. This does not guarantee an improvement, since the field strength decreases with geocentric distance; the relevant optimization target is therefore the integrated magnetic exposure, $\mathcal E_B=\int_{\rm occ}dt\,\mathcal I_B^2(t)$, rather than the occultation duration alone.

Several limitations remain, for ejemplo T89 is a controlled large-scale representation rather than an event-specific magnetospheric reconstruction; the plasmaspheric density is treated semi-empirically; the propagated trajectory is a realizability example and not an optimized mission orbit and the detector response is representative, while the actual background must be determined for the specific mission and observing configuration. In addition, because the expected counts are of order unity, a mission-specific analysis should replace the Asimov benchmark by the exact spatial, spectral, and temporal likelihood appropriate to the instrument.

A positive signal would also provide directional information about the solar axion source. Primakoff axions originate predominantly in the solar interior, and the spatial/spectral axion image can in principle be inverted to constrain radial solar properties \cite{Hoof2023}. Lunar occultation is particularly natural for such a measurement because ordinary solar x rays are suppressed while the weakly interacting messenger remains visible through downstream conversion.

In summary, a physically propagated low-Earth orbit can realize a lunar occultation geometry with a coherent geomagnetic lever arm of order $10^2~\mathrm{T\,m}$ for several minutes. The present August 2026 example gives $\Delta t_{\rm occ}\simeq470$ s and $\mathcal{I}_{B,\mathrm{rms}}=76.9~\mathrm{T\,m}$ for a central $4'$ ROI containing $97.4\%$ of the 3--15 keV Primakoff flux. The associated detector benchmark for one single ocultation exposure does not improve existing limits, but it establishes lunar occultation as an independent and experimentally structured geomagnetic helioscope concept.

\textit{Data and code availability.---}
The numerical code and derived data used to reproduce the results
presented in this work are publicly available in Ref.~\cite{SuarezFontanellaCode2026}.

\section*{Acknowledgments}
I thank Dr. German Sborlini for insightful discussions and valuable feedback that helped improve this manuscript.

\bibliography{references}

\end{document}


\title{Supplemental Material for ``Light Shining Through the Moon:
A Lunar-Occultation Helioscope for Solar Axions''}

\author{D. Suárez-Fontanella}
\affiliation{Department of Fundamental Physics and IUFFyM, University of Salamanca,
	Plaza de la Merced SN, E–37008 Salamanca, Spain}

\maketitle

\section{Vector geomagnetic conversion}

For every detector position the propagation direction is the apparent solar direction, not the detector--Moon direction. The Moon enters only through the occultation mask. With $\hat{\ell}$ denoting the solar line-of-sight unit vector,
\begin{equation}
\mathbf B_\perp(s) =\mathbf B(s)-[\mathbf B(s)\cdot\hat{\ell}]\,\hat{\ell}.
\end{equation}
A fixed orthonormal transverse basis $\{\hat{\mathbf e}_1,\hat{\mathbf e}_2\}$ is constructed for each line of sight and the signed components are
\begin{equation}
B_i(s)=\mathbf B(s)\cdot\hat{\mathbf e}_i,\qquad i=1,2.
\end{equation}
The phase-resolved integrals are
\begin{equation}
\mathcal I_i(E,\ma)=\int ds\,B_i(s)e^{i\phi(s)},
\,
\phi(s)=\int_0^s ds' \frac{\ma^2-\omega_{\rm pl}^2(s')}{2E},
\end{equation}
and
\begin{equation}
\Pag=\frac{\ga^2}{4} \left(|\mathcal I_1|^2+|\mathcal I_2|^2\right).
\end{equation}
This treatment retains cancellations due to sign changes or rotations of the transverse field.

In the coherent limit,
\begin{equation}
\IB=\left[ \left(\int_0^{s_{\max}} B_1(s)\,ds\right)^2+ \left(\int_0^{s_{\max}} B_2(s)\,ds\right)^2 \right]^{1/2}.
\end{equation}
Here $s=0$ denotes the spacecraft position and $s_{\max}$ is the first intersection of the sunward line of sight with the geocentric sphere $r=15R_\oplus$. The same limits are used throughout the conversion calculation.

For comparison, we define the scalar diagnostic
\begin{equation}
\mathcal I_{|B|} = \int_0^{s_{\max}}|\mathbf B_\perp(s)|\,ds .
\end{equation}
This quantity discards cancellations caused by changes in the transverse field direction and is therefore not used in the physical conversion probability.

The field is modeled as the sum of IGRF-14 \cite{Beggan2026} and the T89 external field \cite{Tsyganenko1989}. The fiducial calculation uses $K_p=2$. At the orbital anchor,
\begin{center}
\begin{tabular}{c c c}
\toprule
$K_p$ & $\mathcal I_{|B|}$ (T m) & $\IB$ (T m)\\
\midrule
0 & 87.165 & 71.989\\
2 & 87.532 & 71.889\\
4 & 87.844 & 71.866\\
\bottomrule
\end{tabular}
\end{center}
showing explicitly the cancellation relative to the scalar-magnitude integral and the small anchor-level variation over this illustrative $K_p$ range.

The conversion code uses
$$
1~{\rm T}=195~{\rm eV}^2,\qquad
1~{\rm m}=5.0677307\times10^6~{\rm eV}^{-1},
$$
so that
\begin{equation}
1~{\rm T\,m}=0.9882075~{\rm GeV}.
\end{equation}
For example, $\IB=96.42~{\rm T\,m}$ and $\ga=10^{-10}~{\rm GeV}^{-1}$ imply the coherent estimate $\Pag\simeq2.27\times10^{-17}$, consistent with the low-mass plateau of the numerical calculation.

\section{Plasma model and numerical phase integration}

The plasma frequency is
\begin{equation}
\omega_{\rm pl}=3.69\times10^{-11}\sqrt{n_e/{\rm cm}^{-3}}\ {\rm eV}.
\end{equation}
For the numerical feasibility calculation we use a smooth semi-empirical profile based on the Carpenter--Anderson equatorial plasmasphere parameterization \cite{CarpenterAnderson1992}. The reference density $n_{\rm pl}=10^{3.9043-0.3145L}\,{\rm cm}^{-3}$ and the $K_p$-dependent plasmapause location $L_{\rm pp}=5.6-0.46K_p$ follow that model, while beyond the plasmapause we adopt a simple $r^{-2}$ continuation to represent the dilute outer magnetosphere. Writing $r$ in Earth radii,
\begin{align}
L_{\rm pp}&={\rm clip}(5.6-0.46K_p,\,2.5,\,6.0),\\
n_{\rm pl}(r)&=10^{3.9043-0.3145r}\ {\rm cm}^{-3},\\
n_{\rm out}(r)&=5\left[\frac{\max(r,1)}{10}\right]^{-2}
{\rm cm}^{-3},\\
w(r)&=\left[1+\exp\!\left(\frac{r-L_{\rm pp}}{0.18}\right)\right]^{-1},\\
n_e(r)&={\rm clip}\!\left[
w\,n_{\rm pl}+(1-w)n_{\rm out},\,10^{-4},\,10^5
\right]{\rm cm}^{-3}.
\end{align}
This profile provides a controlled phase contribution for the present feasibility calculation.

A direct quadrature becomes inefficient when the phase varies rapidly. The magnetic and density profiles are therefore interpolated onto a finer path grid whose local phase advance does not exceed a prescribed value. For $K_p=2$ the convergence check gives
\begin{center}
\begin{tabular}{c c c c}
\toprule
$\ma$ (eV) & $\Delta\phi_{\max}$ & $\Pag$ & change to $0.05$ rad\\
\midrule
$3\times10^{-5}$ & 0.10 & $3.62447\times10^{-18}$ & 0.0223\%\\
$3\times10^{-5}$ & 0.05 & $3.62528\times10^{-18}$ & --\\
$1\times10^{-4}$ & 0.10 & $3.55352\times10^{-20}$ & 0.1246\%\\
$1\times10^{-4}$ & 0.05 & $3.55795\times10^{-20}$ & --\\
\bottomrule
\end{tabular}
\end{center}
which supports retaining the plotted mass range through $10^{-4}$ eV.

\begin{figure}[t]
\centering
\includegraphics[width=\columnwidth]{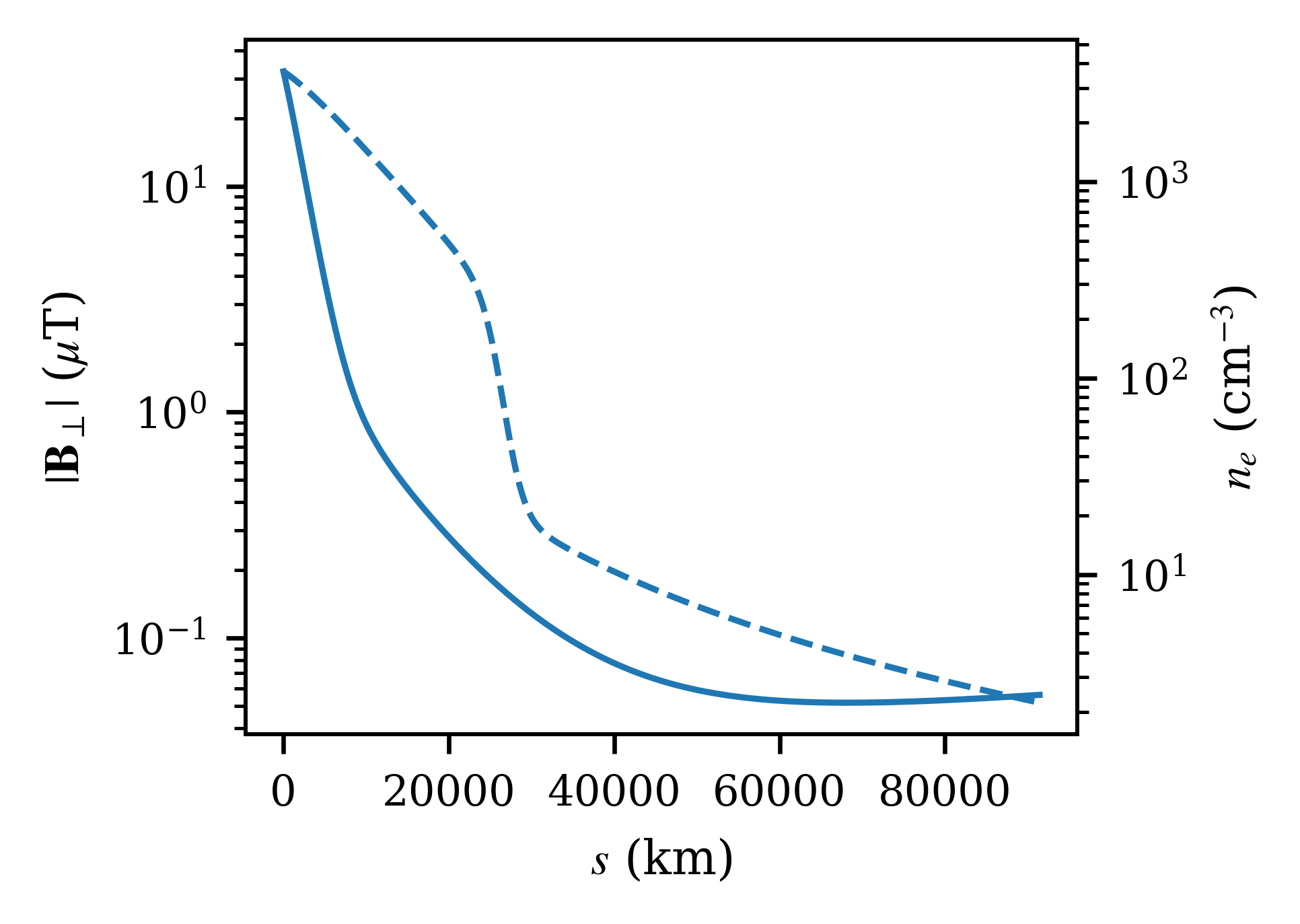}
\caption{Representative transverse-field magnitude and electron-density profile along a sunward conversion path. The magnetic magnitude is shown only as a diagnostic; the conversion probability uses the two signed transverse components separately.}
\label{fig:supp_los}
\end{figure}

\section{Solar-axion region of interest}

The signal region is a circular aperture of radius $\theta_{\rm ROI}=4.0'$ centered on the apparent solar center. At the fiducial epoch,
\begin{equation}
\alpha_\odot=15.7779',
\qquad
\rho_{\rm ROI}=\frac{\theta_{\rm ROI}}{\alpha_\odot}=0.253519.
\end{equation}
The full-Sun Primakoff spectrum is multiplied by an energy-dependent aperture fraction,
\begin{equation}
f_{\rm ROI}(E)=\frac{\displaystyle\int_0^{\rho_{\rm ROI}}d\rho\,\frac{d^2\Phi_a^P}{dE\,d\rho}}{\displaystyle\int_0^1d\rho\,\frac{d^2\Phi_a^P}{dE\,d\rho}}.
\end{equation}
We evaluate the projected Primakoff emission using the B16-AGSS09 standard solar model \cite{Vinyoles2017} and the SolarAxionFlux implementation used in Ref.~\cite{Hoof2023}. The resulting 121-point table spans 3--15 keV in 0.1-keV steps. Selected values are
\begin{center}
\begin{tabular}{c c}
\toprule
$E$ (keV) & $f_{\rm ROI}(E)$\\
\midrule
3 & 0.91955462\\
5 & 0.98279225\\
8 & 0.99827838\\
10 & 0.99962093\\
15 & 0.99999223\\
\bottomrule
\end{tabular}
\end{center}
and the Primakoff-spectrum-weighted fraction over 3--15 keV is
\begin{equation}
\overline f_{\rm ROI}^{\,3-15}=0.973734.
\end{equation}
No uniform-surface-brightness approximation is used.

\begin{figure}[t]
\centering
\includegraphics[width=\columnwidth]{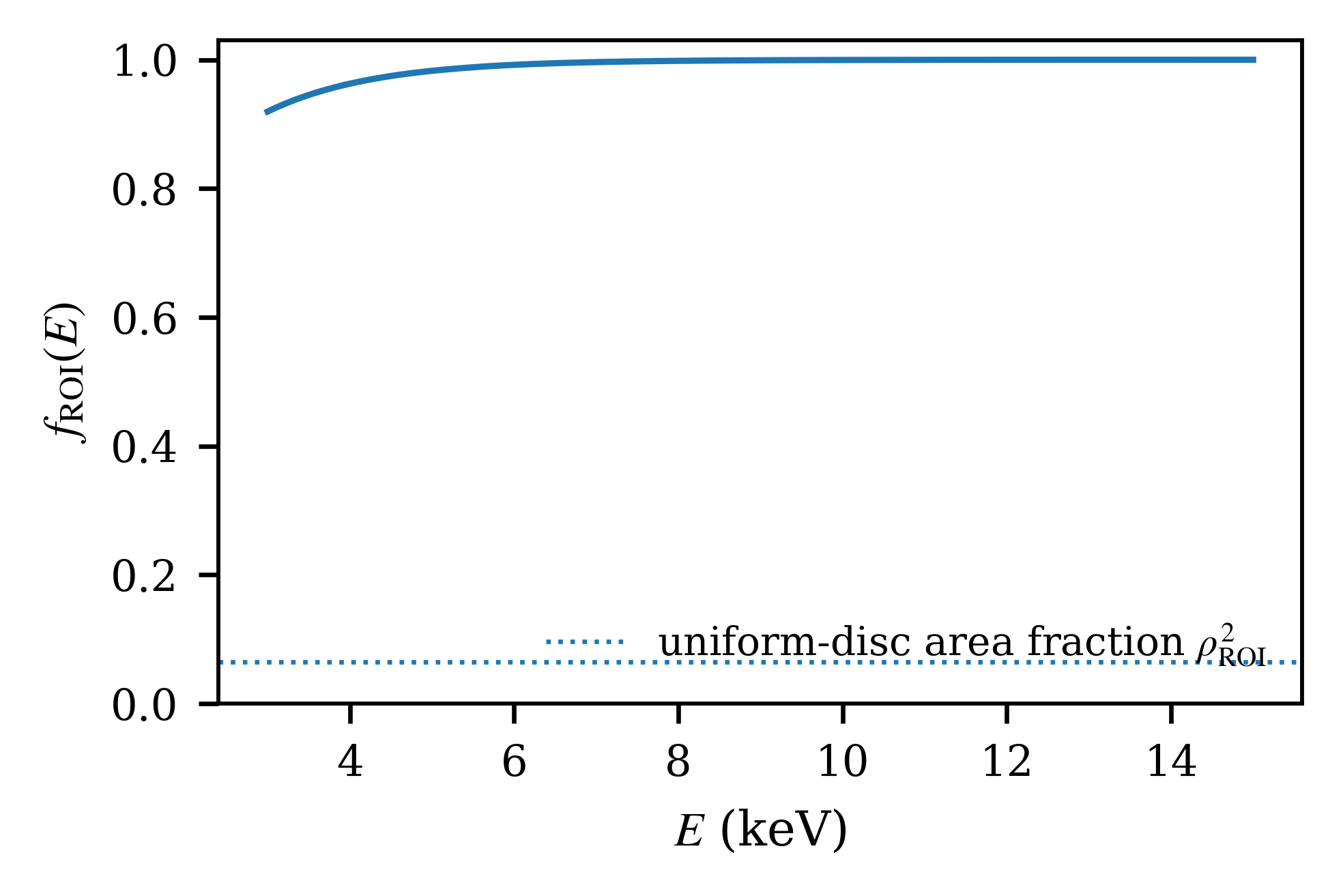}
\caption{Energy-dependent fraction of the B16-AGSS09 Primakoff solar-axion flux contained inside the adopted $4'$ aperture. The uniform-disc area fraction is shown only as a geometric reference.}
\label{fig:supp_roi}
\end{figure}

\section{Propagated circular-orbit realization}

The orbit is anchored at 
$$
t_{\rm occ}={\rm 2026\mbox{-}08\mbox{-}12~17:45:54~UTC}
$$

with geocentric inertial state
\begin{align}
\mathbf r_0&=(-3374.271152,\,-299.272668,\,6082.106660)\ {\rm km},\\
\mathbf v_0&=(-6.584683,\,-0.584012,\,-3.681830)\ {\rm km\,s^{-1}}.
\end{align}
These vectors define the demonstrative circular polar orbit
\begin{align}
r_{\rm orb}&=6961.845~{\rm km},\\
v_{\rm orb}&=7.5667~{\rm km\,s^{-1}},\\
T_{\rm orb}&=96.349~{\rm min},\\
i&=90^\circ.
\end{align}
The anchor is at nominal WGS84 altitude 600 km; subtracting the fixed $6371.2$-km reference radius used elsewhere in the calculation gives approximately 590.6 km, hence the small difference between quoted altitude conventions.

For a circular two-body orbit,
\begin{equation}
\mathbf r(t)=r_{\rm orb} \left[ \hat{\mathbf r}_0\cos(nt) +\hat{\mathbf v}_0\sin(nt) \right],
\qquad
n=\sqrt{\frac{GM_\oplus}{r_{\rm orb}^3}}.
\end{equation}
At each propagated position the apparent Sun--Moon geometry and complete sunward magnetic path are recomputed. A sample is accepted only when
\begin{equation}
\delta_{\odot M}(t)+\theta_{\rm ROI}<\alpha_M(t)
\end{equation}
and the line of sight is clear of the Earth.

The valid interval runs from $-250$ to $+220$ s relative to the anchor, corresponding to 470 s. During it,
\begin{align}
	\mathcal{I}_{B,\min} &=69.147~{\rm T\,m},\\
	\mathcal{I}_{B,\max} &=96.420~{\rm T\,m},\\
	\langle \IB \rangle_{\rm sample}
	&=76.611~{\rm T\,m}.
\end{align}
The signal-weighted quantities used in the Letter are
\begin{align}
\EB&=\int_{\rm occ}dt\,\IB^2(t) =2.778552\times10^6~({\rm T\,m})^2{\rm s},\\
\mathcal{I}_{B,\mathrm{rms}}&=76.888~{\rm T\,m}.
\end{align}
At the anchor, the independently constructed line of sight gives $\IB=71.904392~{\rm T\,m}$ compared with $71.888896~{\rm T\,m}$ in the original fiducial geometry, a relative difference of $0.0216\%$.

\begin{figure}[t]
\centering
\includegraphics[width=\columnwidth]{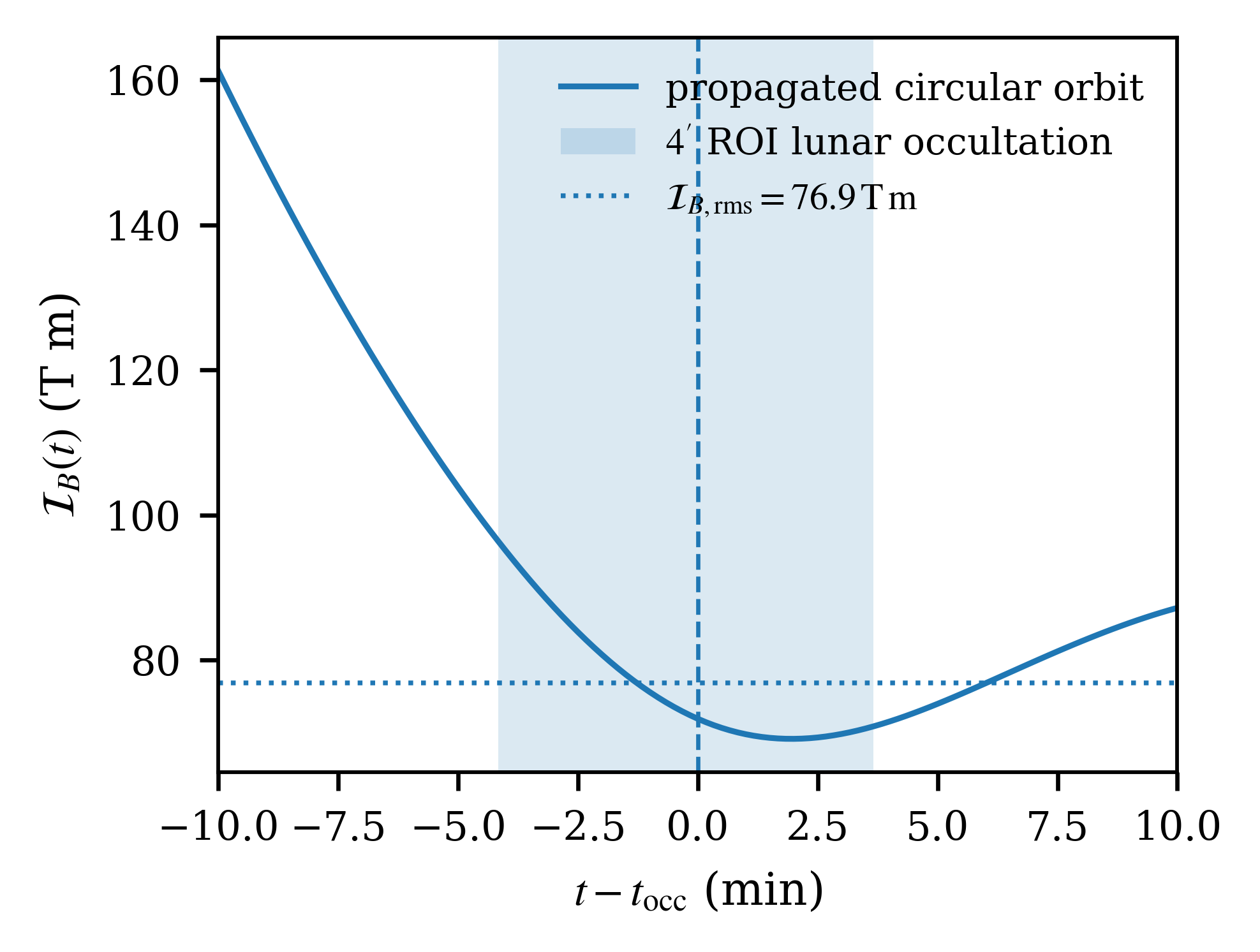}
\caption{Coherent magnetic lever arm along the propagated circular polar orbit. The highlighted interval corresponds to samples for which the $4'$ solar ROI is fully hidden by the Moon and the sunward line of sight remains clear of the Earth.}
\label{fig:supp_orbit}
\end{figure}

\section{Detector benchmark and expected counts}

The sensitivity curve in the Letter uses a instrument-scale benchmark. The combined focusing-telescope effective area is obtained by linear interpolation of
\begin{center}
\begin{tabular}{c c|c c}
\toprule
$E$ (keV) & $A_{\rm eff}$ (cm$^2$) &
$E$ (keV) & $A_{\rm eff}$ (cm$^2$)\\
\midrule
3  & 100 & 10   & 900\\
4  & 350 & 12   & 820\\
5  & 600 & 15   & 700\\
6  & 750 & 20   & 560\\
8  & 850 & 30   & 380\\
   &     & 50   & 180\\
   &     & 78.4 & 5\\
\bottomrule
\end{tabular}
\end{center}
although only the 3--15 keV interval enters the present analysis.

The circular $4'$ aperture has area
\begin{equation}
\Omega_{\rm ROI}= \pi(4')^2=50.265~{\rm arcmin}^2.
\end{equation}
The adopted aperture-integrated background rate over 3--15 keV is
\begin{equation}
R_B=2.0\times10^{-3}~{\rm count\,s^{-1}}.
\end{equation}
For numerical integration its spectral shape is taken to be
\begin{equation}
b(E)=C\left(0.7+0.3\frac{E}{10~{\rm keV}}\right),
\end{equation}
where $C$ is fixed so that $\int_{3\,{\rm keV}}^{15\,{\rm keV}}b(E)dE=R_B$. The 470-s event therefore contains
\begin{equation}
B=R_B\Delta t_{\rm occ}=0.9400
\end{equation}
expected background counts.

For each mass,
\begin{align}
S_{\rm ref}(\ma) = \int_{\rm occ}dt\int_{3}^{15}dE\, A_{\rm eff}(E) \frac{d\Phi_a^{\rm ROI}}{dE} \\
\times \Pag(E,\ma,t;\ga^{\rm ref}), \nonumber
\end{align}
with $\ga^{\rm ref}=10^{-10}~{\rm GeV}^{-1}$. The final numerical run samples the physical occultation every 10 s and uses 60 energy points over the 3--15 keV interval. At low mass,
\begin{equation}
S_{\rm ref}=0.8717068.
\end{equation}
The corresponding signal for another coupling is
\begin{equation}
S(\ga)=S_{\rm ref} \left(\frac{\ga}{\ga^{\rm ref}}\right)^4.
\end{equation}

For the expected background-only benchmark we use
\begin{equation}
q_{A} =2\left[ S-B\ln\left(1+\frac{S}{B}\right) \right]
\end{equation}
and solve $q_{A}=2.71$ for the one-sided 95\% C.L. sensitivity. This gives
\begin{equation}
\ga^{95}=1.31442\times10^{-10}~{\rm GeV}^{-1}
\end{equation}
in the low-mass plateau. Representative values extracted from the final mass scan are
\begin{center}
\begin{tabular}{c c}
\toprule
$\ma$ (eV) & $\ga^{95}$ (GeV$^{-1}$)\\
\midrule
$1.0\times10^{-8}$ & $1.314\times10^{-10}$\\
$9.4\times10^{-7}$ & $1.314\times10^{-10}$\\
$1.06\times10^{-5}$ & $1.290\times10^{-10}$\\
$3.26\times10^{-5}$ & $1.542\times10^{-10}$\\
$1.0\times10^{-4}$ & $3.996\times10^{-10}$\\
\bottomrule
\end{tabular}
\end{center}

With both $S$ and $B$ of order unity for a single event, the Asimov construction serves as a reproducible estimate of the expected sensitivity. A mission-level analysis should instead employ the exact Poisson or profile likelihood appropriate to the selected instrument

\begin{figure}[t]
\centering
\includegraphics[width=\columnwidth]{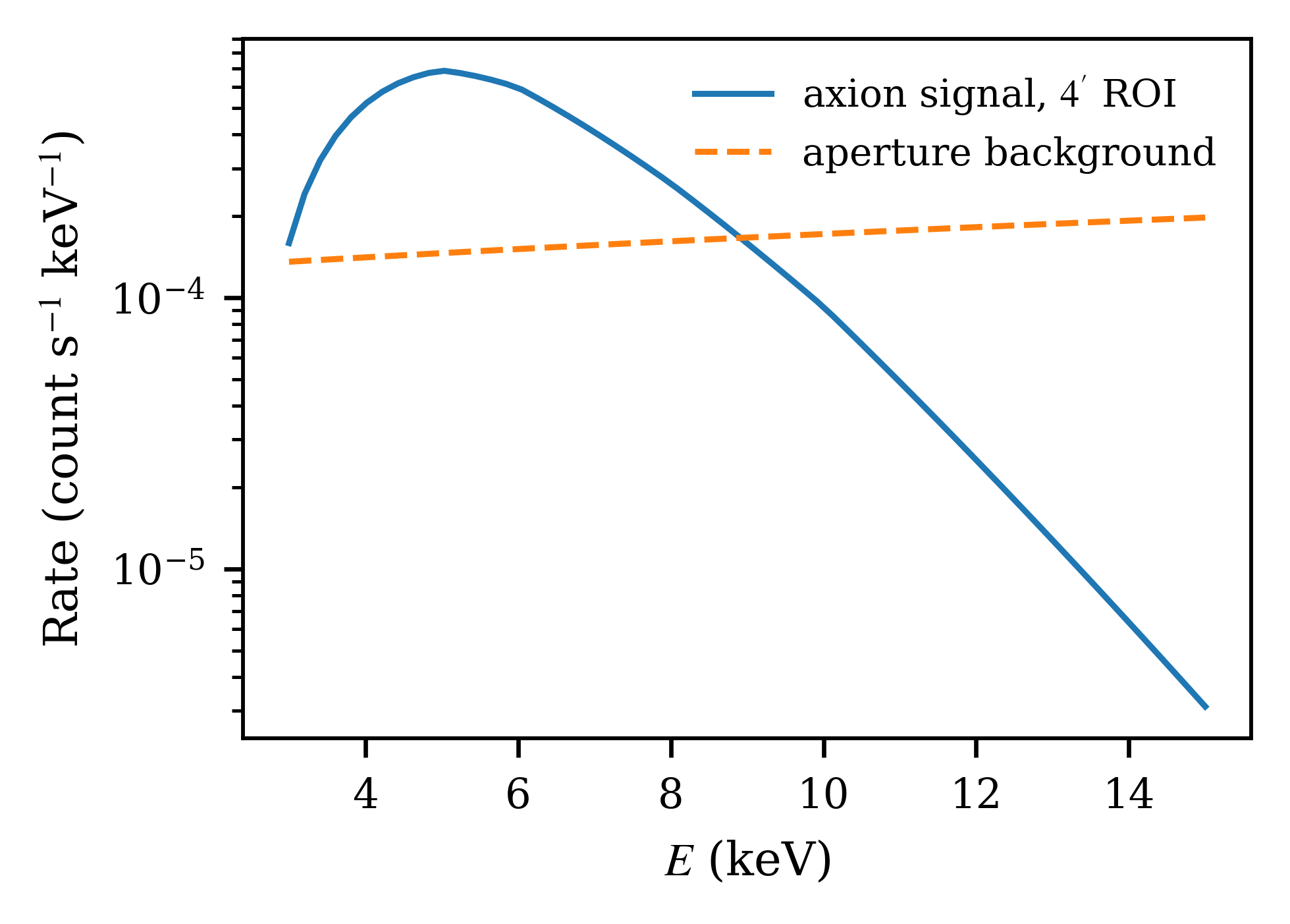}
\caption{Benchmark regenerated-axion signal and aperture-integrated detector background spectra over the 3--15 keV analysis band.}
\label{fig:supp_spectrum}
\end{figure}

\section{Relation to previous geomagnetic helioscope studies}

The present proposal builds directly on the geomagnetic conversion mechanism identified in Ref.~\cite{DavoudiaslHuber2006}. The detailed GECOSAX study \cite{DavoudiaslHuber2008} propagated realistic satellite orbits, included the nonuniform geomagnetic field and atmospheric x-ray absorption, and used a measured Suzaku background. For $10^3~{\rm cm^2}$ detector area and roughly $10^6$ s of exposure, it projected a $2\sigma$ reach $\ga<(4.7$--$6.6)\times10^{-11}~{\rm GeV}^{-1}$ for $\ma<10^{-4}$ eV. The lunar configuration considered here has only a few-hundred-second exposure per event and is consequently weaker in raw
single-event sensitivity. Its distinguishing element is the upstream lunar occultor, which suppresses direct solar x rays before geomagnetic regeneration and defines a spatially and temporally localized search region.

\section{Scope of the feasibility calculation}

The propagated August trajectory demonstrates dynamical accessibility but is not an optimized mission orbit. A mission-level study must additionally use: (i) a specified spacecraft orbit and long-term event accessibility; (ii) event-specific solar-wind and magnetospheric inputs; (iii) calibrated effective-area, redistribution, point-spread-function and background products for the chosen x-ray telescope; (iv) actual Sun/Moon avoidance, thermal and stray-light constraints; and (v) an exact spatial, spectral and temporal likelihood. These refinements affect the instrument forecast but do not change the basic conversion geometry
demonstrated here.

\bibliography{references}